\documentclass{aa}  
\usepackage{natbib}
\usepackage{graphicx}
\usepackage{txfonts}
\usepackage{hyperref} 
\def\j1419{\mbox{J1419$+5246$}}      
\defcitealias{Zhang2025ApJS..280....3Z}{Z25}                    

\begin{document}

   \title{Indirect evidence of the $2175\,\mathring{\mathrm{A}}$ extinction bump within the dusty torus of SDSS J141945.50+524648.0}
%   \title{Indirect evidence for a circumnuclear origin of the $2175\,\mathring{\mathrm{A}}$ extinction bump in SDSS J1419+5246}

   \author{Ze Li\inst{1} 
        \and Gaoyang Chen\inst{1} 
        \and Qifan Cui\inst{1}
        \and Zheng Cai\inst{2}
        \and Jianzhen Chen\inst{1}\thanks{Corresponding author: jzchen@shnu.edu.cn}
        \and Zhijian Luo\inst{1}
		\and Chenggang Shu\inst{1}        
		\and Fengwu Sun\inst{3}
		\and Hubing Xiao\inst{1}
		\and Shaohua Zhang\inst{1}\thanks{Corresponding author: zhangshaohua@shnu.edu.cn}         
		}

   \institute{Shanghai Key Lab for Astrophysics, Shanghai Normal University, Shanghai 200234, People’s Republic of China
   \and Department of Astronomy, Tsinghua University, Beijing 100084, People’s Republic of China
   \and Center for Astrophysics $|$ Harvard \& Smithsonian, 60 Garden St., Cambridge, MA 02138, USA}

   \date{Received xx xx, xxxx; accepted xx xx, xxxx}

% \abstract{}{}{}{}{}
% 5 {} token are mandatory
  \abstract
  {We present a multi-wavelength study of the quasar SDSS J141945.50$+$524648.0 ($z=1.1599$), a member of the newly identified population of quasar-associated $2175\,\mathring{\mathrm{A}}$ dust absorbers.	  
  Utilizing high-quality JWST/NIRCam observations from the SAPPHIRES survey and archival data, we analyze the prominent $2175\,\mathring{\mathrm{A}}$ extinction bump detected in this object.   
  The extinction bump parameters are highly consistent with those of the Milky Way extinction curves, implying similar dust properties. 
  Spectral analysis reveals a $\mathrm{Mg\,II}$ absorption doublet near the systemic velocity. A joint fit to the $\mathrm{Mg\,II}$ and $\mathrm{Fe\,II}$ absorption lines favors a partial-covering model, 
  with the absorber fully covering the compact continuum source while only partially covering the broad-line emission region, yielding a covering factor of $C_{f,\rm BEL}=0.11^{+0.12}_{-0.07}$.
  Together with the small velocity offset ($\Delta v\approx96~{\rm km\,s^{-1}}$), these results favor an intrinsic absorber rather than a simple intervening system.
  Modeling the infrared spectral energy distribution indicates a dusty torus viewing angle of $\theta \approx 60^{+7}_{-8}$ deg, while the double-peaked $\mathrm{Pa,\alpha}$ emission profile suggests an accretion-disk inclination of $i = 63.7^{+10.9}_{-7.8}$ deg.
  The agreement between these independent geometric constraints suggests a "rim-penetrating" line of sight through the dusty torus.
  This configuration provides a self-consistent explanation for the infrared emission, continuum extinction, and associated $2175\,\mathring{\mathrm{A}}$ extinction bump, although a contribution from the host-galaxy ISM cannot be ruled out.
  If the absorber is indeed associated with the torus, the carbonaceous carriers of the $2175\,\mathring{\mathrm{A}}$ feature may survive the intense AGN radiation field through localized shielding in optically thick dusty clumps.
  These findings highlight that the detection of such extinction features may be closely tied to the viewing geometry and dust distribution, offering new insights into dust survival and destruction mechanisms in quasar environments.}


   \maketitle

\section{Introduction}
The $2175\,\mathring{\mathrm{A}}$ extinction bump is the most prominent ultraviolet (UV) feature in the Milky Way (MW) extinction curve \citep{Gordon2024JOSS....9.7023G}, yet it is generally weak or absent in the Small and Large Magellanic Clouds (SMC and LMC)  \citep{Gordon2003ApJ...594..279G, Gordon2024ApJ...970...51G}. 
Although the exact carriers of this feature have remained debated since its discovery \citep{Stecher1965ApJ...142.1683S}, carbonaceous materials—such as polycyclic aromatic hydrocarbons (PAHs), hydrogenated T-carbon (HTC), or graphite grains—are considered the leading candidates for the $2175\,\mathring{\mathrm{A}}$ extinction bump carriers. \citep[e.g.,][]{LiDraine2001ApJ...554..778L, Ma2020MNRAS.497.2190M}.

Detecting this feature at high redshift is challenging due to both observational limitations and the evolving physical properties of dust.
Observationally, the lack of standard background stars precludes the traditional ``stellar pair" technique employed in the MW, while the intrinsically diverse and poorly constrained continua of quasars or galaxies introduce significant degeneracies between extinction and the source spectrum  \citep[e.g.,][]{Maiolino2001A&A...365...37M,Jiang2011ApJ...732..110J,Wang2012ApJ...760...42W,Ma2017MNRAS.472.2196M}. 
Physically, the dust composition and size distribution at high redshift may differ substantially from those in the local Universe. 
The interstellar medium (ISM) in many distant galaxies tends to exhibit lower metallicities and harder radiation fields than those in nearby systems \citep[e.g.,][]{Madau2014ARA&A..52..415M,Maiolino2019A&ARv..27....3M}. 
Such conditions may hinder the formation or survival of the carbonaceous dust carriers believed to be responsible for the $2175\,\mathring{\mathrm{A}}$ extinction feature \citep{Voit1992MNRAS.258..841V,Draine2003ARA&A..41..241D}.
Recent theoretical and observational studies further suggest that the abundance and processing of carbonaceous grains are strongly influenced by metallicity, radiation hardness, and grain growth in the ISM \citep[e.g.,][]{Madden2006A&A...446..877M,Galliano2018ARA&A..56..673G,Witstok2023Natur.621..267W}.

Despite these hurdles, the $2175\,\mathring{\mathrm{A}}$ extinction bump has been successfully identified in numerous intervening absorption systems, such as $\mathrm{Mg\,II}$, $\mathrm{Zn\,II}$, $\mathrm{Ca\,II}$, or $\mathrm{C\,I}$ absorbers \citep[e.g.,][]{Wang2004ApJ...609..589W, Srianand2008MNRAS.391L..69S, Jiang2011ApJ...732..110J, Ma2015MNRAS.454.1751M, Ma2018MNRAS.474.4870M, Fang2023MNRAS.518.5590F, Ge2024MNRAS.531..387G, Wang2025arXiv251013863W}, and damped Ly$\alpha$ systems (DLAs) \citep[e.g.,][]{Wucknitz2003AA...405..445W,Junkkarinen2004ApJ...614..658J, Wang2012ApJ...760...42W, Pan2017ApJ...835..218P}.
Intervening systems are absorbers located along the line of sight but physically unrelated to the background quasars, whereas quasar-associated systems have redshifts close to the quasar systemic redshift, typically corresponding to velocity offsets of $v_{\rm off} \le 5000$ km s$^{-1}$.
Unlike quasar-associated absorbers, the dust in intervening systems is generally not directly exposed to the intense UV and X-ray radiation of the quasar central engine, residing instead in the more typical ISM of foreground galaxies \citep[e.g.,][]{York2006MNRAS.367..945Y, 2008MNRAS.385.1053M}. 
Quasar-associated systems, in contrast, serve as unique probes of dust physics in environments exposed to intense AGN radiation fields, typically characterized by high ionization parameters ($U \sim 10^{-2}-1$; e.g., \citealt{1999ARA&A..37..487H,Arav2013MNRAS.436.3286A,Zhang2015ApJ...815..113Z, Zhang2018AJ....156....4Z, 2019MNRAS.487.5041H, Pan2019ApJ...883..173P}),
although the exact origin of the dust extinction in these systems remains uncertain. Yet detections of such carriers in these environments have remained exceptionally rare.
The intense UV radiation from the central engine is expected to destroy the dust carriers, and indeed, only $\sim18$ such absorber candidates \citep[e.g.,][]{Gaskell2007arXiv0711.1013G, Jiang2011ApJ...732..110J, Zhang2015ApJ...802...92Z, Zhang2022A&A...663A..63Z, Pan2017ApJ...835..218P, Shi2020ChA&A..44..196S, Zhou2022ApJ...941..111Z}
were known prior to the study by S. Zhang et al. (2025; hereafter \citetalias{Zhang2025ApJS..280....3Z}), 
who identified a large sample of 843 quasar-associated absorbers in the final Sloan Digital Sky Survey IV (SDSS-IV) quasar catalog from Data Release 16 (DR16; \citealt{Lyke2020ApJS}), expanding the known population by nearly 50-fold.
The existence of such a substantial population challenges previous assumptions about dust destruction in harsh radiation environments, suggesting that $2175\,\mathring{\mathrm{A}}$ carriers can survive and even thrive in the immediate vicinity of quasars.
We note, however, that although the redshifts of the $2175\,\mathring{\mathrm{A}}$ absorbers match those of the background quasars, this does not necessarily imply that the responsible dust grains reside within the active galactic nucleus (AGN) environment. 
Instead, the absorption may originate in the ISM of the quasar host galaxy or its galactic halo \citep[e.g.,][]{York2006MNRAS.367..945Y, 2008MNRAS.385.1053M}. 
Considering these alternative scenarios is therefore essential for correctly interpreting the origin of the $2175\,\mathring{\mathrm{A}}$ extinction feature.

In this work, we focus on SDSS J141945.50$+$524648.0 (hereafter \j1419), currently the only object in this sample with James Webb Space Telescope / Near-Infrared Camera (JWST/NIRCam) coverage. 
Its location within the Extended Groth Strip (EGS) region coincides with a wealth of JWST observations, making \j1419 a prime candidate for characterizing this new population in the context of a rich multi-wavelength environment.
By employing multi-component spectral energy distribution (SED) modeling and spectroscopic analysis, 
we constrain the composition and location of the dust responsible for this unique absorption, thus providing new insights into dust physics in extreme AGN environments.

\section{Observations and Reduction}

\begin{table*}[ht]
	\centering
	\tiny
	\caption[]{Photometric and spectroscopic data of \j1419 \label{tab:table_obs}}
	\begin{tabular}{cccccc}
		\hline \hline
		Bandpass&Measurement&Unit&Telescope/Instrument&Aperture Type&Reference\\
		\hline
		u		&21.861$\pm$0.154 &asinh mag	&SDSS			&	modeled datum	&	1 \\
		g		&21.455$\pm$0.049 &asinh mag	&SDSS			&	modeled datum	&	1 \\
		r		&20.787$\pm$0.042 &asinh mag	&SDSS			&	modeled datum	&	1 \\
		i		&20.636$\pm$0.052 &asinh mag	&SDSS			&	modeled datum	&	1 \\
		z		&20.355$\pm$0.136 &asinh mag	&SDSS			&	modeled datum	&	1 \\
		W1		&16.356$\pm$0.053 &mag			&WISE			&from fitting to map&	2 \\
		W2		&15.183$\pm$0.062 &mag			&WISE			&from fitting to map&	2 \\
		W3		&11.877$\pm$0.194 &mag			&WISE			&from fitting to map&	2 \\
		W4		& 9.346$\pm$0.499 &mag			&WISE			&from fitting to map&	2 \\
		NUV 		&22.920$\pm$0.070 &AB mag	&GALEX			&	fixed aperture	&	3 \\
		F606W 		&21.060$\pm$0.001 &AB mag	&HST/ACS		&	fixed aperture	&	3 \\
		F814W 		&20.540$\pm$0.001 &AB mag	&HST/ACS		&	fixed aperture	&	3 \\
		Ks			&19.430$\pm$0.020 &AB mag	& Hale/WIRC		&	fixed aperture	&	3 \\
		3.6  $\mu$m &18.810$\pm$0.040 &AB mag	&Spitzer/IRAC	&	fixed aperture	&	3 \\
		4.5  $\mu$m	&18.530$\pm$0.040 &AB mag	&Spitzer/IRAC	&	fixed aperture	&	3 \\
		5.8  $\mu$m	&18.140$\pm$0.030 &AB mag	&Spitzer/IRAC	&	fixed aperture	&	3 \\
		8.0  $\mu$m	&17.570$\pm$0.040 &AB mag	&Spitzer/IRAC	&	fixed aperture	&	3 \\
		24   $\mu$m	&1363.0$\pm$5.778 &$\mu$Jy	&Spitzer/MIPS	&	modelled datum 	&	4 \\
		F090W		&24.433$\pm$0.024 &$\mu$Jy	&JWST/NIRCam  	&	fixed aperture	& 	6 \\
		F200W		&54.089$\pm$0.009 &$\mu$Jy	&JWST/NIRCam  	&	fixed aperture	& 	6 \\
		F444W		&142.207$\pm$0.048&$\mu$Jy	&JWST/NIRCam 	&	fixed aperture	& 	6 \\
		\hline
		$[3569.44,10334.7]$ \AA	&	  &			&BOSS			&					&	5 \\
		$[2.4,5.0]~\mu$m		&	  &			&JWST/NIRCam 	&					&	6 \\ %
		\hline
	\end{tabular}
	\tablebib{  (1)~\cite{Lyke2020ApJS}; 
		(2) \cite{Cutri2013wise}; 
		(3) \cite{Nandra2015}; 
		(4) \cite{SpitzerIRSA};  
		(5) \cite{Ahumada2020ApJS..249....3A}; 
		(6) This work}
\end{table*}

\j1419 is a quasar previously identified in the SDSS Data Release 12 quasar catalog \citep{2017A&A...597A..79P}, with its systemic redshift updated to $z = 1.1599$.\footnote{The systemic redshift was recalibrated using the $\mathrm{Pa}\alpha$ emission line captured in the JWST/NIRCam infrared spectrum.}	
Its optical spectrum was first taken with the 2.5 m SDSS telescope on 2013 June 12, with a total exposure time of 5405 seconds. The one-dimensional spectrum used in this work was retrieved from the SDSS DR16 \citep{Ahumada2020ApJS..249....3A}. 
The BOSS spectrographs provide spectra with a resolving power of $R \approx 1560$–$2650$ over a wavelength range of approximately 3600–10400 \AA\ \citep{Smee2013AJ....146...32S, Dawson2013AJ....145...10D}.
To examine the properties of the broadband SED, photometric data spanning from the near-ultraviolet (NUV) to the mid-infrared (MIR) were compiled from the Galaxy Evolution Explorer (GALEX; \citealt{Morrissey2007ApJS..173..682M}), SDSS \citep{York2000AJ....120.1579Y}, and Wide-field Infrared Survey Explorer (WISE; \citealt{Wright2010AJ....140.1868W}). 
In addition, \cite{Nandra2015} provided optical and infrared photometry of \j1419 from the Hubble Space Telescope / Advanced Camera for Surveys (HST/ACS), 
the Wide-field Infrared Camera (WIRC) on the 200-inch Hale Telescope, 
and the Infrared Array Camera (IRAC) on the Spitzer Space Telescope
in the Rainbow database \citep{Barro2011ApJS..193...30B,Ramos2012MNRAS.419..687R}. 
We further supplemented the Ks-band magnitude with data from \cite{SpitzerIRSA}.

\j1419 is also located within the field of view of the Slitless Areal Pure-Parallel HIgh-Redshift Emission Survey (SAPPHIRES; \citealt{Sun2025arXiv250315587S}), a JWST Cycle-3 Treasury large survey using the NIRCam wide-field slitless spectroscopic (WFSS) mode in pure parallel. 
The SAPPHIRES observations for \j1419 provide a total of 23.4 hours of data, taken on 2025 February 27-28 and March 1. These data include imaging in three filters (F090W, F200W, and F444W) and long-wavelength (LW; $2.4-5.0\, \mu\mathrm{m}$) spectroscopy. 
The primary dataset is drawn from program 6434 (observations 231, 232, and 443; \citealt{Egami2024jwst.prop.6434E}).
All NIRCam imaging data were reprocessed using the latest stable version of the JWST Calibration Pipeline (version 1.16.1; \citealt{Bushouse2024zndo..14153298B}), together with the Calibration Reference Data System (CRDS) pipeline mapping 1321.
Source detection and aperture photometry were carried out on the images with \texttt{SExtractor} \citep{Bertin1996A&AS..117..393B} with a fixed aperture of $1.0\arcsec$. 
The NIRCam LW grism spectrum was processed using the public available routine presented by \cite{Sun2023ApJ...953...53S}, including standard Stage-1 reduction with the JWST calibration pipeline and various customized steps, e.g., flat-fielding, modeled background subtraction, astrometric calibration, 2D spectral extraction and flux/wavelength calibration. The 1D spectrum was extracted using a standard boxcar method with an aperture height of 0."3, and the aperture flux loss have been corrected using the spatial profile of the spectral trace.

All the photometric and spectral data from the aforementioned surveys, supplemented by the additional observations, are summarized in Table~\ref{tab:table_obs}. 
The SDSS photometric and spectroscopic observation for \j1419 were obtained with a temporal separation of approximately 10 years. We noted significant discrepancies between the observed photometry and the synthetic magnitudes derived from the spectrum ($0.10, 0.21, 0.31,$ and $0.26$ mag in the $g, r, i,$ and $z$ bands, respectively).
Although such offsets could, in principle, be attributed to intrinsic variability \citep[e.g.,][]{Guo2016ApJ...822...26G}, an analysis of the long-term light curves from the  Palomar Transient Factory (PTF; \citealp{Law2009PASP..121.1395L}) and Zwicky Transient Facility (ZTF; \citealp{Masci2019PASP..131a8003M}) data shows that the source has remained %remarkably stable over the past two decades. 
broadly stable over the past two decades within the statistical uncertainty of the individual measurements (typically 0.2 - 0.5 mag).
This suggests that the observed discrepancies are unlikely to be driven by long-term variability, and may instead arise from systematic differences between photometric and spectroscopic calibrations.
To ensure a self-consistent SED analysis, we applied a flux correction to the SDSS spectrum, anchoring it to stable photometric measurements, following the method introduced in \citetalias{Zhang2025ApJS..280....3Z}.
In combining multi-instrument datasets with different extraction apertures, we performed careful consistency check. Given the predominantly point-source-dominated nature of J1419+5246, as confirmed by high-resolution HST/ACS and JWST/NIRCam imaging, the aperture effects are expected to be minimal after standard corrections.
Our visual inspection of JWST/NIRCam images confirms that no  nearby sources bright enough to significantly contaminate the larger WISE apertures. 
Furthermore, the good agreement among flux measurements in the SDSS $z$, HST/ACS F814W, and JWST F090W, between the Hale Ks and JWST F200W bands,
and between the WISE W2, Spitzer/IRAC 4.5$\mu$m and JWST F444W bands,
provides a robust basis for our cross-calibration, ensuring that the resulting SED reliably represents the intrinsic emission of the quasar.

\section{Analysis and Results} 

In this section, we perform a comprehensive analysis of the spectral and photometric characteristics of \j1419, aiming to uncover its dust extinction properties, infrared emission components, and the kinematic structure of its broad emission lines.

\subsection{$2175\,\mathring{\mathrm{A}}$ Dust Extinction}

\begin{figure*}[ht]
%	\figurenum{1}
	\centering
	\includegraphics[width=0.9\hsize]{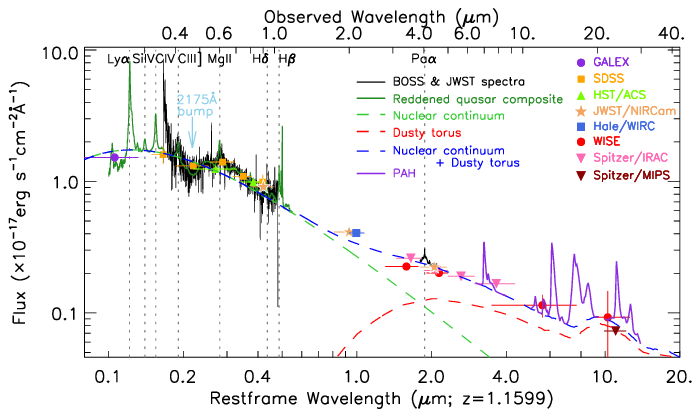}
	\caption{Broadband SED of \j1419 from NUV to MIR. 
	Photometric measurements from different instruments are shown with distinct symbols and colors, as indicated in the legend. 
	The recalibrated BOSS spectrum and the JWST/NIRCam WFSS spectrum are plotted as black lines. 
	The nuclear continuum and the dusty torus emission are shown as green and red dashed curves, and their combined model is overplotted as a blue dashed curve. 
	The forest-green curve represents the reddened quasar composite spectrum, while the purple features indicate the PAH emission components. 
	The sky-blue arrow marks the position of the $2175\,\mathring{\mathrm{A}}$ extinction bump.}
	\label{fig1}
\end{figure*}

In Figure \ref{fig1}, the photometrically recalibrated BOSS spectrum corrected for foreground Galactic extinction using the updated reddening map of \citet{Schlafly2011ApJ...737..103S} and the extinction law characterization of \citet{Fitzpatrick2019ApJ...886..108F}, is shown in black.
Based on this spectrum, \citetalias{Zhang2025ApJS..280....3Z} reported the presence of a quasar-associated $2175\,\mathring{\mathrm{A}}$\ dust absorber, which was identified by comparing the observed spectrum with the quasar composite \citep{VandenBerk2001AJ....122..549V,Glikman2006ApJ...640..579G} and deriving the extinction curve from the resulting spectral ratio.
In this work, we adopt the parameterized extinction curve directly from the measurements in \citetalias{Zhang2025ApJS..280....3Z}.  
The five parameters characterizing the parameterized extinction curve \citep{Fitzpatrick2007ApJ...663..320F} are $c_1 = 8.09 \pm 0.03$ mag, {$c_2 = 0.21 \pm 0.01$ mag, $c_3 = 0.23 \pm 0.05$ mag, \mbox{$x_0 = 4.62 \pm 0.04$ $\rm \mu m^{-1}$}, and $\gamma = 0.89 \pm 0.11$ $\rm \mu m^{-1}$ (see Section 3 and Table 1 of \citetalias{Zhang2025ApJS..280....3Z}).
The forest green curve in the figure represents the reddened quasar composite, which well reproduces the BOSS spectrum. In both spectra, the broad and deep absorption bump 
is centered at $\sim 2162\,\mathring{\mathrm{A}}$ in the quasar rest frame, which corresponds to the famous $2175\,\mathring{\mathrm{A}}$ extinction bump.
The strength of the bump can be described by the area under the Drude profile, $A_{\rm bump} = \int_{0}^{\infty} c_3 D(x,x_{0}, \gamma)\, \mathrm{d}x = \pi c_3 / 2\gamma$, and then $A_{\rm bump} = 0.41 \pm 0.10$ ${\rm mag}\,\mu{\rm m}^{-1}$, corresponding to a significance of $3.55\sigma$.
Details of the optical spectral fitting and the identification of the  $2175\,\mathring{\mathrm{A}}$ dust absorber can be found in Figure 1.625. of \citetalias{Zhang2025ApJS..280....3Z}.

\subsection{Infrared Emission from the Torus}

Figure \ref{fig1} presents the multi-band photometric data as orange squares, with error bars representing the band widths and flux errors.
A reddened UV/optical continuum (green dashed curve) is overlaid. This continuum uses an intrinsic slope of \mbox{$\alpha_\lambda = -1.54$}, which is adopted from the quasar composite of \citet{VandenBerk2001AJ....122..549V}. The reddening is applied using only the power-law terms ($c_1$ and $c_2$) of the parameterized extinction curve. This approach successfully traces the continuum windows \citep{VandenBerk2001AJ....122..549V} of \j1419, with the notable exception of the $2175\,\mathring{\mathrm{A}}$ bump region.

Extrapolating the power-law component into the infrared regime, the photometric data are found to be significantly higher than the continuum, indicating a pronounced infrared excess in the SED of \j1419. 
This excess emission suggests the presence of the hot dusty torus--a familiar component of the AGN structure. 
We modeled this emission excess by adding a torus component (red dashed curve), and the combined continuum and dust model is displayed in blue.
The torus component is derived from the self-consistent AGN torus template library presented in \citet{Siebenmorgen2015A&A...583A.120S}. 
This torus model describes the circumnuclear dust as a two-phase medium (a clumpy medium and/or a homogeneous disk). 
The resulting torus SED is parameterized by a set of basic AGN and torus geometrical and physical parameters.

Since the torus component is modeled using a library of discrete emission templates rather than a continuous parametric function, the model space is inherently restricted to a fixed grid of physical parameters. The best-fit for \j1419 was determined through a systematic search across the template library. 
For each discrete template, we determined the optimal linear scaling factor using least-squares minimization via the MPFIT algorithm \citep{Markwardt2009ASPC..411..251M}. 
The final model, corresponding to the template \mbox{``R0300\_Vc077\_Ac0135\_Ad0300.th5"}, was selected by identifying the global minimum $\chi^2$ across the entire grid. 
When selecting the model, the reddening was also taken into account.
This model yields a viewing angle of \mbox{$\theta =60^\circ$} measured from the polar axis, an inner radius for the dusty torus of \mbox{$r_{\rm in} = 300 \times 10^{15} \text{ cm}$}, and a cloud volume filling factor of \mbox{$\eta =7.7\%$}. 
Furthermore, the $V$-band optical depths of the individual clouds and the disk midplane are \mbox{$\tau_{V,{\rm cl}} = 13.5$} and \mbox{$\tau_{V,{\rm mid}} = 300.0$}, respectively.
It should be noted that the five parameters in the template library are discretized on a grid within their respective ranges; the detailed setup is summarized in Table 1 of \citet{Siebenmorgen2015A&A...583A.120S}. 
For the viewing angle, the library adopts nine discrete values, among which the two closest to $60^\circ$ are $52^\circ$ and $67^\circ$.
Accordingly, a pragmatic estimate of the viewing-angle uncertainty is approximately $\pm 7-8^\circ$.

We note that the photometric fluxes in the Spitzer/IRAC \mbox{8.0  $\mu$m}, WISE W3 and W4  bands remain slightly elevated relative to  the combined continuum from the nucleus and the dusty torus (blue curve). 
This excess emission may be partly attributed to PAH emission. 
To illustrate this, we overlaid the PAH emission features from the star-forming region, the Orion Bar \citep{peeters2021spectroscopic}, which spans the range from $3\,\mu\text{m}$ to $14\,\mu\text{m}$ and is shown as the purple solid curves in the figure.
The Mid-Infrared Instrument (MIRI) on board JWST offers both Low-Resolution Spectroscopy \citep[LRS;][]{Kendrew2015PASP..127..623K} and Medium-Resolution Spectroscopy \citep[MRS;][]{Wells2015PASP..127..646W} capabilities. 
The LRS covers the spectral range of $5-12~\mu\text{m}$, while the MRS spans from $5~\mu\text{m}$ to $28~\mu\text{m}$. 
These spectral windows are well-suited to cover the potential PAH emission features in \j1419. Consequently, future MIRI observations would allow us to verify the hypothesized contribution of PAHs to the observed MIR excess.

\subsection{Double-peaked Profile of $\mathrm{Pa\,\alpha}$  Emission Line}

\begin{figure}[ht]
	%	\figurenum{2}
	\centering
	\includegraphics[width=\hsize]{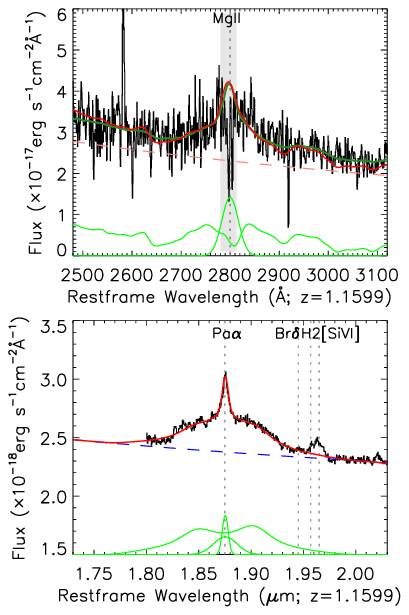}
	\caption{
		Spectra of the $\mathrm{Mg\,II}$ and $\mathrm{Pa\,\alpha}$ regions with best-fit models.  
		Top: dust-extinction-corrected BOSS spectrum of \j1419 (black) and quasar composite (forest green). Pink shows nuclear continuum; green shows broadened UV $\mathrm{Fe\,II}$ template and Gaussian $\mathrm{Mg\,II}$ emission line; red shows total model.  
		The shaded region denotes the region of $\mathrm{Mg\,II}$ absorption troughs.
		Bottom: JWST/NIRCam WFSS spectrum of \j1419 (black). Blue shows combined continuum model; green shows narrow and broad Gaussian components plus double-peaked disk component; red shows total model.}
	\label{fig2}
\end{figure}

The spectroscopic data for \j1419, utilizing the UV coverage from the BOSS and the near-infrared (NIR) coverage from the JWST/NIRCam observations, span a broad wavelength range that captures the $\mathrm{Mg\,II}$ and $\mathrm{Pa\,\alpha}$ emission lines in the rest frame, respectively. This unique multi-wavelength coverage enables a comprehensive study of its broad emission line (BEL) properties.

The top panel of Figure \ref{fig2} presents the dust-extinction–corrected BOSS spectrum of \j1419 (black curve), which shows excellent agreement with the quasar composite (forest green curve), as expected.
Following the standard approach described in \citet{Zhang2010ApJ...714..367Z}, the spectral modeling is performed over a broad rest-frame wavelength range of $2100 \text{ to } 3100 \text{ \AA}$. This range is chosen to ensure a reliable decomposition of the continuum and the complex $\mathrm{Fe\,II}$ features. During the fitting process, spectral regions contaminated by $\mathrm{Mg\,II}$ absorption troughs are carefully masked to prevent them from biasing the emission-line profile reconstruction; the masked region are indicated by the shaded area in Figure \ref{fig2}.

Within this range, the spectrum is modeled using three primary components:
1) the power-law nuclear continuum (pink dashed line), characterized by a spectral slope of $\alpha_{\lambda} = -1.54$; 
2) the broadened UV $\mathrm{Fe\,II}$ emission template \citep{Vestergaard2001ApJS..134....1V} (green curve); 
and 3) the $\mathrm{Mg\,II}$ BEL, modeled as a single Gaussian component (also green) with a Full Width at Half Maximum ($\text{FWHM}$) of $4240\pm 387 \text{ km/s}$. 
The total model is shown as the red curve. 
Applying the relation of the black hole mass to the $\mathrm{Mg\,II}$ line width and the extinction-free luminosity \citep{Wang2009ApJ...707.1334W}, we estimate that \mbox{$M\rm_{BH}=(1.87\pm0.58)\times10^8\rm\, M_{\sun}$}. Given the quasar luminosity of \mbox{$L\rm_{bol}=(1.26\pm0.12)\times10^{45}\, erg/s$}, the corresponding Eddington ratio is $L_{\rm bol}/L_{\rm Edd} \approx 0.05$.

The bottom panel of Figure \ref{fig2} presents the JWST/NIRCam WFSS spectrum (black curve), which crucially covers the rest-frame $\mathrm{Pa\,\alpha}$ emission line of \j1419. Overlaid on this observational data is the total combined continuum model (blue dashed line) derived from the accretion disk and dusty torus, as detailed in Section 3.2. In contrast to the $\mathrm{Mg\,II}$ emission line, the $\mathrm{Pa\,\alpha}$ emission profile is complex, exhibiting a narrow emission component alongside an underlying, exceptionally broad emission profile reaching a $\text{FWHM}$ of $14,000 \text{km/s}$.

To decompose the $\mathrm{Pa\alpha}$ emission profile, we employed a three-component model—a standard approach widely adopted in the literature for capturing complex broad-line profiles
\citep[e.g.,][]{Bergmann2017ApJ...835..236S,Zhang2019ApJ...877...33Z, Zhang2019MNRAS.490.1738Z,Santos2023ApJ...953L...3D}.The model includes:
1) A narrow Gaussian component ($\text{FWHM} = 1063 \pm 35 \text{ km/s}$) representing the emission from the narrow line region (NLR).
2) A broad Gaussian component accounting for the emission from the classical broad line region (BLR). To maintain kinematic consistency across different wavebands and reduce the number of free parameters, we fixed its profile in velocity space to those of the $\mathrm{Mg\,II}$ BEL, following our previous methodology \citep[e.g.,][]{Zhang2019ApJ...877...33Z}. 
This component represents the traditional, likely more isotropic or turbulent BLR gas.
3) A double-peaked component representing the broad emission originating directly from the rotating accretion disk.

Here, the double-peaked disk component is modeled using the elliptical accretion disk model developed by \cite{Eracleous1995ApJ...438..610E}. 
While complex features such as bars \citep[e.g.,][]{Chakrabarti1994ApJ...434..518C, Karas2001PASJ...53..189K}, warped structures \citep[e.g.,][]{Hartnoll2000MNRAS.317..880H}, or hotspots \citep[e.g.,][]{Zheng1991ApJ...381..418Z, Lewis2010ApJS..187..416L} primarily influence the subtle, small-scale profile variations, the elliptical model effectively captures the large-scale body structure of the emission regions. 
Although established decades ago, this model remains a robust and widely used benchmark for characterizing double-peaked broad lines in AGNs \citep[e.g.,][]{Strateva2003AJ....126.1720S, Zhang2025A&A...699A.245Z, Marsango2026MNRAS.546ag182M}.
This model is defined by seven free parameters: the inner boundary $r_{\text{in}}$, the outer boundary $r_{\text{out}}$, the inclination angle $i$, the radius-dependent line emissivity $f(r) \propto r^{-q}$, the local broadening velocity $\sigma$, the eccentricity $e$, and the apocenter direction angle $\phi_0$.
The MPFIT algorithm yields the following best-fit parameters for \j1419: $r_{\text{in}} = 227\pm31$ $\text{R}_g$, $r_{\text{out}} = 4297\pm175$ $\text{R}_g$, $i = 63.7^{+10.9}_{-7.8}$ deg, $q=1.74\pm0.21$, $\sigma=1505\pm243$ $\text{km/s}$, $e=0.11\pm0.04$, and $\phi=125.1\pm17.2$ deg (where $\text{R}_g$ denotes the gravitational radius).
To ensure the reliability of the fitting results and address potential parameter degeneracies, we conducted the fitting with various initial parameter sets. The consistent convergence to the reported values suggests that the best-fit model represents a stable solution. Furthermore, by fixing the kinematic profile of the broad Gaussian component and only allowing its normalization to vary, we minimized the artificial coupling between the Gaussian and the double-peaked disk components.
Notably, the resulting inclination angle is highly consistent with the viewing angle ($\theta \approx 60^{+7}_{-8}$ deg) independently inferred from the torus model in Section 3.2. 
Moreover, if future follow-up NIR spectroscopy of \j1419 detects the Balmer emission lines, it would provide a crucial and independent verification of the fitted parameters for the disk component.

\begin{figure*}[ht]
	%	\figurenum{3}
	\centering
	\includegraphics[width=\hsize]{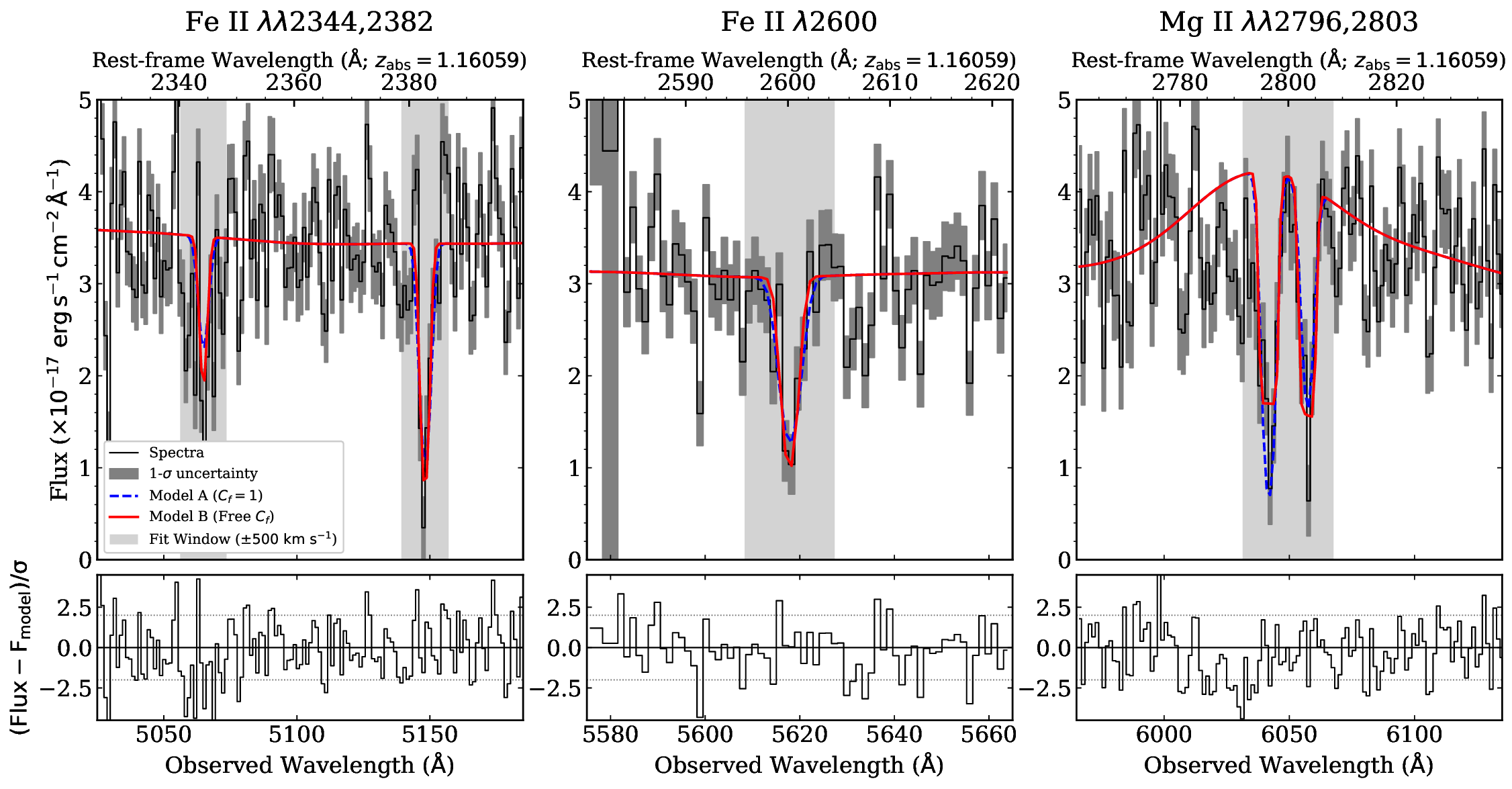}
	\caption{Joint Voigt-profile fitting of the $\mathrm{Mg\,II}$ $\lambda\lambda 2796, 2803$ and $\mathrm{Fe\,II}$ $\lambda\lambda 2344, 2382, 2600$ absorption line systems. 
	In the upper panels, the dust-extinction-corrected BOSS spectrum (also shown in the top panel of Figure \ref{fig2}) is presented as a black step line, with the $1$-$\sigma$ uncertainty represented by the gray shaded region. 
	The light gray vertical bands indicate the velocity windows 
	($\pm500\,\mathrm{km\,s^{-1}}$ around the line centers) included in the joint fitting. 
	Two models are overplotted: Model A assumes full coverage of both the continuum and BEL regions ($C_{f,\rm cont}=C_{f,\rm BEL}=1$; blue dashed line), whereas Model B assumes full coverage of the continuum source but a free covering factor for the BEL region ($C_{f,\rm cont}=1$, free $C_{f,\rm BEL}$; red solid line). The continuum and BEL contributions are treated separately in the partial-covering model.
	The bottom subpanels display the normalized fit residuals for Model B in units of standard deviation, $(Flux - F_{\mathrm{model}})/\sigma$, with dotted horizontal lines indicating the $\pm 2\,\sigma$ levels. 
	The bottom and top axes display the observed wavelength and the corresponding rest-frame wavelength at $z_{\mathrm{abs}}$, respectively. 
	Model B is statistically preferred ($\Delta\mathrm{BIC}=11.27$), yielding a BEL-region covering factor of $C_{f,\rm BEL}=0.11_{-0.07}^{+0.12}$, indicating that the intrinsic absorber only partially obscures the background quasar emission.}
	\label{fig3}
\end{figure*}

We examine the $\mathrm{Mg\,II}$ spectral region, where a distinct absorption doublet is identified on the broad emission-line profile.
To characterize the absorption features, we adopted a joint Voigt-profile fitting framework.
This framework simultaneously models the $\mathrm{Mg\,II}$ absorption doublet ($\lambda\lambda2796,2803$) and three associated $\mathrm{Fe\,II}$ transitions ($\lambda\lambda2344,2382,2600$) within a velocity window of $\pm500~\mathrm{km\,s^{-1}}$ around their respective line centers. 
Since these low-ionization species are expected to arise from the same absorbing gas, we constrain them to share a common absorption redshift ($z_{\rm abs}$) and Doppler parameter ($b$), assuming that turbulent or macroscopic velocity dispersion dominates over thermal broadening.

To assess the effects of line saturation and the spatial distribution of the background emission regions, we considered two competing absorption models with different assumptions for the coverage of the continuum-emitting source and the BEL region. 
The background quasar emission is decomposed into the accretion-disk continuum ($F_{\rm cont}$) and BEL emission ($F_{\rm BEL}$), including the fitted $\mathrm{Mg\,II}$ and $\mathrm{Fe\,II}$ emission components shown in Figure~\ref{fig2}.

Model A assumes that the absorbing gas completely covers both the continuum source and the BEL region, with fixed covering factors \mbox{$C_{f,\rm cont}=C_{f,\rm BEL}=1.0$}.
Model B adopts a partial-covering geometry, with full coverage of the continuum source ($C_{f,\rm cont}=1.0$) and a free BEL covering factor ($C_{f,\rm BEL}$).
%The BEL-region covering factor ($C_{f,\rm BEL}$) is treated as a free parameter during the fitting process. 
The observed absorption profile is modeled by separately considering the continuum and BEL contributions:
\mbox{$F_{\lambda}^{\rm model} =F_{\lambda}^{\rm cont}e^{-\tau_{\lambda}}+F_{\lambda}^{\rm BEL}\left[(1-C_{f,\rm BEL})+C_{f,\rm BEL}e^{-\tau_{\lambda}}\right]$,}
where $\tau_{\lambda}$ represents the optical depth of the corresponding absorption transition.

Although $\mathrm{Mg\,II}$ and $\mathrm{Fe\,II}$ absorption arise from the same low-ionization absorber and share the same BEL covering factor in the model, their residual fluxes are not identical because their underlying emission backgrounds differ.
The $\mathrm{Mg\,II}$ doublet is superimposed on the broad $\mathrm{Mg\,II}$ emission profile with additional $\mathrm{Fe\,II}$ emission contribution, and the uncovered fraction of these emission components produces residual flux in the absorption trough. In contrast, $\mathrm{Fe\,II}$ absorption is observed against the continuum and local $\mathrm{Fe\,II}$ emission background, whose relative contributions vary with wavelength.

Model selection was performed using the Bayesian Information Criterion (BIC). Model A yields a total $\chi^2=291.23$ for 65 degrees of freedom ($\mathrm{BIC}=308.17$), whereas Model B provides a significantly better fit, yielding $\chi^2=275.72$ for 64 degrees of freedom ($\mathrm{BIC}=296.90$).  
Following \citet{Kass1995}, $\Delta\mathrm{BIC} > 10$ is generally considered strong evidence in favor of the model with lower BIC.
With $\Delta\mathrm{BIC}=11.27$, our results provide very strong support for the partial-covering model over the full-covering model.

To evaluate parameter uncertainties and degeneracies, we performed a Markov Chain Monte Carlo (MCMC) analysis using the \texttt{emcee} package,
adopting flat physical priors on all fit parameters: \mbox{$\log (N/\mathrm{cm}^{-2}) \in [11.0, 16.0]$}, \mbox{$b \in [10, 200]\,\mathrm{km\,s^{-1}}$}, and \mbox{$C_{f,\rm BEL} \in [0, 1]$}.  
The MCMC analysis was performed to characterize parameter uncertainties and degeneracies.
Owing to the moderate saturation of the $\mathrm{Mg\,II}$ doublet, the Doppler parameter and column density become strongly correlated, producing a non-Gaussian posterior distribution. 
The posterior distribution shows the expected saturation degeneracy, with lower Doppler parameters compensated by higher column densities. 
The MCMC analysis yields \mbox{$b = 109.88_{-18.04}^{+20.86}~\mathrm{km\,s^{-1}}$}, \mbox{$\log(N_{\mathrm{Mg\,II}}/\mathrm{cm}^{-2}) = 15.13_{-0.38}^{+0.56}$}, \mbox{$\log(N_{\mathrm{Fe\,II}}/\mathrm{cm}^{-2}) = 14.41_{-0.12}^{+0.10}$}, and \mbox{$C_{f,\rm BEL} = 0.11_{-0.07}^{+0.12}$}. 
The low $C_{f,\rm BEL}$ indicates that only a small fraction of the BEL emission is covered, %while the compact continuum source remains fully covered, 
implying a small effective projected covering area relative to the BEL region.
Furthermore, the measured Doppler parameter ($b \approx 110\,\mathrm{km\,s^{-1}}$) is unlikely to be dominated by thermal broadening. 
If interpreted purely as thermal motion, this value would correspond to a gas temperature of $T\sim1.5\times10^6$ K for $\mathrm{Mg\,II}$, at which Mg$^+$ would be largely ionized and unable to remain abundant. 
Therefore, the observed velocity width likely represents the integrated macroscopic velocity dispersion of multiple unresolved absorption components, such as turbulent motions or complex kinematic structures in the absorbing gas.

The preferred Model B places the absorption system at a redshift of $z_{\mathrm{abs}}=1.16059_{-0.00005}^{+0.00006}$, corresponding to a small velocity offset of $\Delta v\approx96~\mathrm{km\,s^{-1}}$ relative to the quasar systemic redshift ($z=1.1599$), which falls well within the typical uncertainty of broad-line systemic redshifts. 
Integrating the MCMC-derived absorption profiles over independently defined velocity intervals yields rest-frame equivalent widths of $W_{\rm r,2796}=2.32\pm0.24~\mathrm{\AA}$ and $W_{\rm r,2803}=2.12\pm0.24~\mathrm{\AA}$ for $\mathrm{Mg\,II}$, and $W_{\rm r,2344}=0.78\pm0.21~\mathrm{\AA}$, $W_{\rm r,2382}=1.55\pm0.22~\mathrm{\AA}$, and $W_{r,2600}=1.41\pm0.26~\mathrm{\AA}$ for $\mathrm{Fe\,II}$.

In summary, the low BEL covering factor and the small velocity offset favor an intrinsic narrow absorption line (NAL) interpretation associated with the quasar environment. Such a small effective covering factor is difficult to reconcile with a simple intervening galaxy-scale absorber scenario.

\section{Discussion and Summary}

\begin{figure*}[ht]
	%	\figurenum{4}
	\centering
	\includegraphics[width=0.9\hsize]{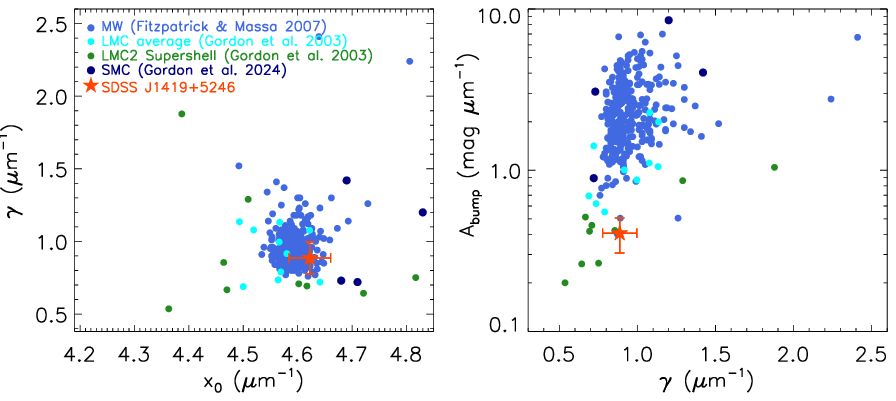}
	\caption{Comparison of bump parameters with $2175\,\mathring{\mathrm{A}}$ bumps in the MW (blue circles), the LMC average (cyan circles), the LMC 2 supershell (green circles), and the SMC (navy circles). The $2175\,\mathring{\mathrm{A}}$ bump in \j1419 is labeled with star.}
	\label{fig4}
\end{figure*}

As discussed in \citetalias{Zhang2025ApJS..280....3Z} and related works, quasar-associated absorbers can generally be divided into two categories: 
absorbers associated with nearby companion galaxies along the line of sight, and absorbers intrinsic to the quasar system itself, such as material located in the circumnuclear environment or the host galaxy. 
For the former, the absorber redshift ($z_{\rm abs}$) is generally lower than the quasar emission redshift ($z_{\rm em}$), although in some rare cases it can exceed $z_{\text{em}}$ \citep[e.g.,][]{1998MNRAS.299..661M}.
Therefore, fixing $z_{\rm abs}=z_{\rm em}$ when modeling the $2175\,\mathring{\mathrm{A}}$ extinction bump can introduce a systematic bias and shift the derived peak position ($x_0$) toward larger values.
These objects most likely correspond to the ``HvNAL'' population presented in \citetalias{Zhang2025ApJS..280....3Z}.

For \j1419, several independent lines of evidence favor an intrinsic origin of the absorber. The Mg II absorption doublet is nearly centered on the broad emission-line profile, with a velocity offset of only $\Delta v\approx96~{\rm km\,s^{-1}}$, strongly suggesting a physical association with the quasar system. In addition, the absorption troughs show non-zero residual flux, indicating partial coverage of the background emission region. Although the exact BLR covering factor depends on the local BLR-to-continuum flux ratio, such partial coverage is commonly observed in intrinsic quasar absorbers.

As shown in Figure~\ref{fig4}, the $2175\,\mathring{\mathrm{A}}$ extinction bump detected in \j1419 exhibits properties remarkably similar to those observed in the Milky Way. 
Here, $A_{\rm bump}$ represents the unnormalized integrated bump area with physical dimensions of ${\rm mag}\,\mu{\rm m}^{-1}$.
To enable a direct comparison with literature studies that report normalized bump strengths \citep[e.g.,][]{Gordon2003ApJ...594..279G, Fitzpatrick2007ApJ...663..320F}, the normalized bump strengths in those works ($A_{\rm bump}/E(B-V)$) are converted to unnormalized values by multiplying by their respective color excesses.
Although the bump strength ($A_{\rm bump}=0.41\pm0.10~{\rm mag}\,\mu{\rm m}^{-1}$) is relatively weak, both the peak position ($x_0=4.62\pm0.04~\mu{\rm m}^{-1}$) and width ($\gamma=0.89\pm0.11~\mu{\rm m}^{-1}$) are consistent with the Galactic average within the measurement uncertainties ($x_0=4.59\pm0.03~\mu{\rm m}^{-1}$, $\gamma=0.95\pm0.16~\mu{\rm m}^{-1}$; \citealp{Fitzpatrick2007ApJ...663..320F}).  
This similarity suggests that the properties of the bump carriers may be comparable to those responsible for the Galactic feature, although detailed constraints on grain composition and size distribution require additional observations.

A key result is the remarkable agreement between the independent geometric constraints derived from NIR analyses.
The best-fitting clumpy torus model for the infrared excess yields a viewing angle of $\theta=60^{+7}_{-8}$ deg, while modeling the double-peaked $\mathrm{Pa\,\alpha}$ emission line with an elliptical accretion-disk model gives an inclination angle of $i=63.7^{+10.9}_{-7.8}$ deg. 
The excellent agreement between these two independent measurements strongly suggests that both the accretion disk and the obscuring torus share a common orientation. 
This aligned geometry naturally places our line of sight near the outer rim of the dusty torus. 
This configuration provides a physically plausible framework that can simultaneously explain the infrared emission, continuum extinction, and the associated $2175\,\AA$ extinction bump.
While this geometry implies that dense nuclear-associated dust plays a major role in producing the extinction, it does not exclude an additional, non-negligible extinction contribution from diffuse dust within the host-galaxy ISM.

If the absorber is associated with the dusty torus, the detection of the $2175\,\mathring{\mathrm{A}}$ extinction bump challenges the expectation that carbonaceous grains cannot survive in the harsh AGN radiation field\citep{Voit1992MNRAS.258..841V,Laor1993ApJ...402..441L}. 
Our best-fitting infrared model adopts a clumpy torus, consistent with modern AGN torus models \citep[e.g.,][]{Krolik1988ApJ...329..702K,Nenkova2008ApJ...685..147N}. 
In this framework, carbonaceous grains may survive if they are embedded within high-optical-depth clumps ($\tau_{\rm V,cl}=13.5$ from our best-fitting clumpy torus model), where the incident radiation field is substantially attenuated \citep{Maiolino2001A&A...365...28M}. 
Such clumpy structures can naturally produce large local optical depths while leaving a significant fraction of the nuclear emission relatively unobscured. 
The survival of carbonaceous grains therefore relies on localized shielding within these optically thick clouds, where the incident radiation field is strongly attenuated.
Therefore, the relatively high value of $\tau_{\rm V,cl}$ inferred from the torus model does not necessarily imply strong reddening of the observed continuum, since the latter is determined primarily by the global distribution and covering fraction of dusty clouds rather than by the optical depth of an individual clump.

This shielding scenario has a physical analogy with broad absorption line (BAL) quasars, where $2175\,\mathring{\mathrm{A}}$ extinction bumps are also found more frequently than in non-BAL quasars \citep[e.g.,][]{Zhang2015ApJ...802...92Z,Zhang2022A&A...663A..63Z,Zhang2025ApJS..280....3Z}. 
In BAL systems, the high-column-density outflowing material associated with BAL troughs can provide an effective shielding environment that may facilitate the survival of carbonaceous grains. 
Although BAL quasars represent a dynamic outflow scenario whereas \j1419 likely probes a relatively static torus-rim configuration, both cases emphasize the importance of geometric shielding and high local opacity in preserving the survival of $2175\,\mathring{\mathrm{A}}$ bump carriers in AGN environments.

Alternatively, the extinction may arise partly or predominantly from dusty material in the host galaxy ISM. We emphasize that associating the Mg II absorber with the dust responsible for the $2175\,\AA$ extinction bump is an assumption rather than a direct observational constraint. A host-ISM origin remains compatible with the Milky Way-like extinction curve, and distinguishing between circumnuclear and host-galaxy dust requires further observations.

Future multi-wavelength observations will be crucial for distinguishing these scenarios.
JWST/MIRI will constrain the warm dust emission and the inner torus component, while future far-infrared (FIR) facilities will probe the cooler dust emission and better constrain the overall torus SED. Complementary submillimeter observations with ALMA, JCMT, and the upcoming Leighton Chajnantor Telescope (LCT; \citealp{Vial2020}) will probe the cold dust component and provide constraints on the dust temperature, mass, and spatial distribution.. 
Higher-resolution optical spectroscopy can resolve the Mg II absorption structure, and future UV spectroscopy targeting high-ionization lines ($\mathrm{C\,IV}$, $\mathrm{Si\,IV}$, and $\mathrm{N\,V}$) and fine-structure transitions ($\mathrm{C\,II^*}$ and $\mathrm{Si\,II^*}$) will constrain the ionization state, density, and location of the absorbing gas

The detection of the $2175\,\mathring{\mathrm{A}}$ extinction bump in J1419 provides a geometry-dependent probe of carbonaceous grain survival near AGNs. Although current evidence provides suggestive support for a circumnuclear interpretation, future observations are required for confirmation. If verified, this result would support the view that AGN extinction curves depend strongly on viewing geometry\citep{Gaskell2007arXiv0711.1013G}, highlighting the role of line-of-sight structure in regulating dust survival and observed extinction properties.

\begin{acknowledgements}
We acknowledge the reviewer for their insightful suggestions that have greatly helped us to present the results better.
This work is supported by the National Natural Science Foundation of China (grant Nos. 12173026, 12141302, 12573009), the Innovation Program of Shanghai Municipal Education Commission (grant No. 2025GDZKZD04), the scientific research grants from the China Manned Space Project (grant Nos. CMS-CSST-2025-A06, CMS-CSST-2025-A07). S.H.Z. acknowledges the support from the Program for Professor of Special Appointment (Eastern Scholar) at Shanghai Institutions of Higher Learning, and from the Shuguang Program (23SG39) of Shanghai Education Development Foundation and Shanghai Municipal Education Commission.
\end{acknowledgements}

\bibliographystyle{aa} % style aa.bst
\bibliography{j1419references} % your references Yourfile.bib

\end{document}